\documentclass[jgrga]{agutex}
\usepackage{graphicx}

\usepackage{wrapfig}

\setkeys{Gin}{draft=false}

\authorrunninghead{{WINTER and BALASUBRAMANIAM}}
\titlerunninghead{{SOLAR FLARE CLASS PREDICTIONS}}
\authoraddr{Corresponding author: Lisa Winter, Space Weather \& Effects Group, Atmospheric and Environmental Research, Superior, CO, USA. (lwinter@aer.com)}

\begin{document}

\lefthead{Winter \& Balasubramaniam}
\righthead{Solar Flare Class Predictions}

\title{Using the Maximum X-ray Flux Ratio and X-ray Background to Predict Solar Flare Class}  

\authors{L.M. Winter\altaffilmark{1} and K. Balasubramaniam\altaffilmark{2}}

\altaffiltext{1}{Atmospheric and Environmental Research, Superior, CO, USA.}
\altaffiltext{2}{Air Force Research Laboratory, Albuquerque, NM, USA.}

\begin{abstract}

We present the discovery of a relationship between the maximum ratio of the flare
flux (namely, 0.5-4 \AA~to the 1-8 \AA~flux) and non-flare background (namely,
the 1-8~\AA~background flux), which clearly separates flares into classes by peak flux level.  
We established this relationship
based on an analysis of the Geostationary Operational Environmental
Satellites (GOES) X-ray observations of $\sim$ 50,000 X, M, C, and B flares derived
from the NOAA/SWPC flares catalog.   Employing a combination of machine
learning techniques (K-nearest neighbors and nearest-centroid algorithms) we
show a separation of the observed parameters for the different peak flaring
energies. This analysis is validated by successfully predicting the flare
classes for 100\% of the X-class flares, 76\% of the M-class flares, 80\% of
the C-class flares and 81\% of the B-class flares for solar cycle 24, based
on the training of the parametric extracts for solar flares in cycles 22-23.
\end{abstract}

\begin{article}
\section{Introduction}
Solar flares release intense amounts of energy into the interplanetary medium.  A statistical concept of how this energy is released is through an avalanche model \citep{1991ApJ...380L..89L,1993ApJ...412..841L,1995ApJ...446L.109L}.  According to common concensus, the creation of a solar flare begins with magnetohydrodynamic instabilities that release energy stored in the local magnetic field lines through an untwisting of the field lines.  Conservation of magnetic energy leads to instabilities in nearby regions through an avalanche process that accelerates energetic particles along the large-scale magnetic field lines.  Soft X-rays in the solar corona, which are the topic of the presented analysis, are emitted from the associated  magnetic loops, which are, in turn, connected to the active region surface magnetic fields, as seen in the photosphere.  In this model, all flares are the result of the same physical processes and the ultimate strength of the flare is related to the cascaded number of reconnection events creating the flare.

While the basic mechanism of solar flares is believed to be understood, the detailed physical processes are still too complex to be modeled in a deterministic way for predictions of when a flare will occur.  This is a challenge, since the space weather effects of the energy release can lead to diverse problems to human technology such as satellite damage or inoperability, ionospheric communication interference, and power grid failures.  Therefore, understanding the solar conditions (e.g., magnetic activity, coronal temperature) that precede and lead to solar flares is of vital importance to enhance current predictions of space weather phenomena. 

Current solar flare prediction models rely upon empirical observations, with many predictions based on tracking the properties of solar active regions (e.g., \citealt{2002SoPh..209..171G,2007SpWea...5.9002B, 2011SpWea...9.4003F, 2013SoPh..283..157A,cite-bala2013}).  Since the magnetic active regions are the source of the magnetic energy released in the flare, this approach is well-justified.  However, other observable properties may also provide a diagnostic for the underlying physical conditions in the solar corona leading to flares.  In particular, we present evidence for two easily observable soft X-ray measurements that diagnose the non-flare magnetic energy and coronal temperature (in Section 2).  The parameter-space of these observables distinguishes properties of solar flares of different classes, as described in Section~\ref{sect-diagram}.  Machine learning techniques are used to build classification models applied to historical flares in Section~4.  Finally, the results from the statistical soft X-ray analysis of solar flares are discussed in Section~5. 


\section{X-ray Flare Data}\label{sect-xraydata}
\begin{figure}
\centering
\includegraphics[width=0.45\linewidth]{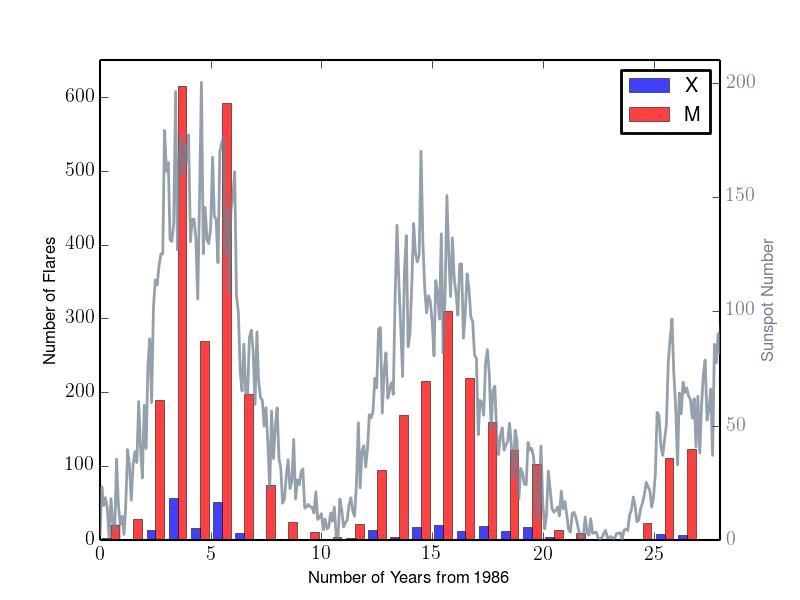}
\includegraphics[width=0.45\linewidth]{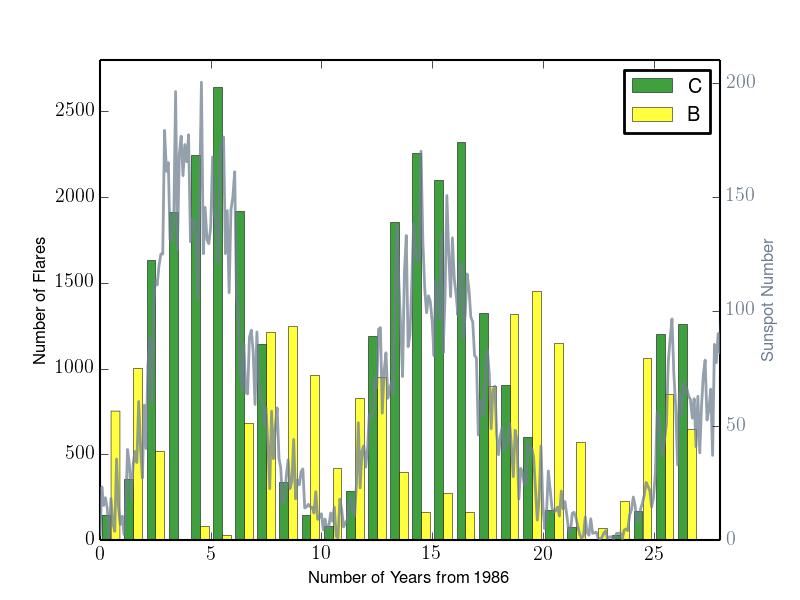}
\caption{The distribution of flares of each type in the NOAA flare list, including observations from 1986 -- July 2014.  Gray lines trace the monthly sunspot number.  The majority of X, M, and C class flares occur close to solar maximum, the maximum in sunspot number.  
}\label{fig-histflares}
\end{figure}

We analyzed the historical X-ray data from NOAA's $GOES$ X-Ray Sensor (XRS).  The $GOES$ observations include X-ray flux measurements averaged over every 1 minute observed in both a short-wavelength and long-wavelength X-ray band (short: 0.5-4\,\AA~and long: 1--8\,\AA) from 1986 -- 2014.  The NOAA flare lists include nearly 50,000 X-ray flares in this timespan which covers solar cycles 22-24.  The flare classifications fall into the following classes based on the peak flux level: X ($> 10^{-4}$\,W\,m$^{-2}$), M ($> 10^{-5}$\,W\,m$^{-2}$), C ($> 10^{-6}$\,W\,m$^{-2}$), and B ($> 10^{-7}$\,W\,m$^{-2}$). 
Figure~\ref{fig-histflares} shows the distribution of the flares of each type by year.  The majority of the stronger flares occur close to solar maximum, indicated in the plots by the maxima in sunspot number (obtained from the Solar Influences Data Analysis Center in Belgium).

The NOAA flare lists include the start, peak, and end time of the flares along with the flare location, if known, and X-ray class.  For our analysis, we use the start and peak time along with the flare class.  The start of a flare is defined as when four consecutive one-minute 1--8\,\AA~flux measurements meet all of the following conditions: (1) All four values $> 10^{-7}$\,W\,m$^{-2}$, (2) each consecutive measurement has a higher flux than the previous measurement, and (3) the last value is $> 1.4\times$ the measurement from three minutes earlier.  

\begin{figure}
\centering
\includegraphics[width=0.8\linewidth]{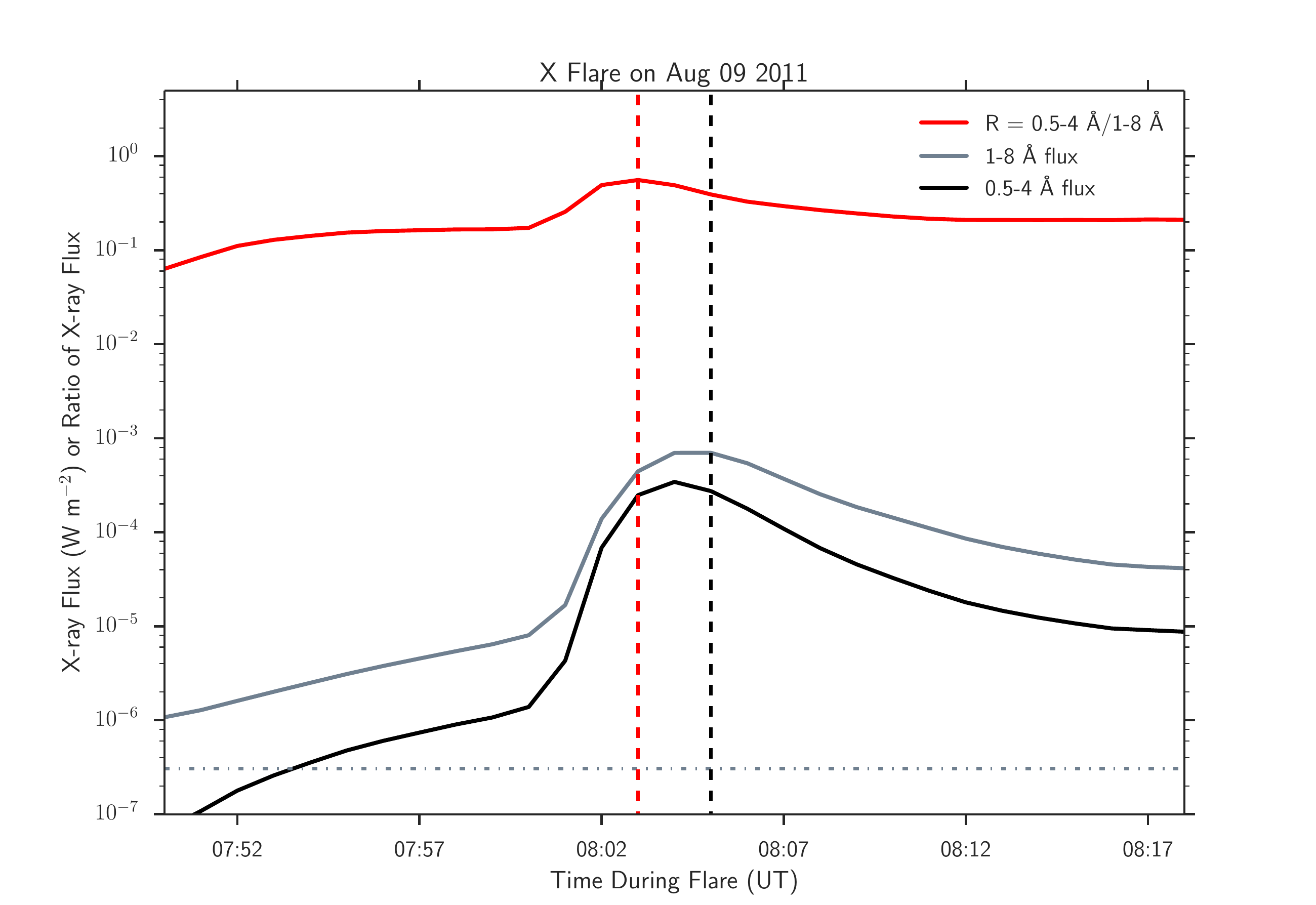}
\caption{For the $\sim$50,000 flares in the NOAA flare lists, occurring from 1986 -- present, we calculated the maximum ratio of the short (0.5-4\,\AA) to long (1--8\,\AA) X-ray bands.  An example is shown for an X-class flare from 2011.  The ratio $R$ (red), long wavelength (gray), and short wavelength (black)  X-ray flare profiles are shown, with the maximum in $R$ marked with a red dashed line and the 1-8\,\AA~flare peak marked with a black dashed line.  The dot-dashed line marks the 1-8\,\AA~non-flare background, $B$.
}\label{fig-ratio}
\end{figure}

Using the downloaded XRS data in both the short and long bands, we measured the long X-ray background flux ($B$) and the ratio of the short to long bands:
\begin{equation} R = \frac{F_{0.5-4~\,\rm{Ang}}}{F_{1-8~\,\rm{Ang}}}.
\end{equation}
  A full description of our method to determine $B$ is included in \citet{2041-8205-793-2-L45}, where we show that the long X-ray background, but not the short X-ray background, varies along with the solar cycle for solar cycles 22-24.  This soft X-ray background variation was also observed for solar cycle 21 by \citet{1988AdSpR...8...67W} and \citet{1994SoPh..152...53A}.
 The background is measured as the minimum 1--8\,\AA~flux in the preceding 24 hours for each 1-minute XRS measurement, following the procedure of
\citet{SWE:SWE20042}.  This background measurement is similar to the X$_{b10}$ index, used in operational forecasting as an integrated irradiance proxy by \citet{Tobiska2006347}. The ratio $R$ is computed from the start time of each flare until the peak, using the dates from the NOAA flare lists.  In a small number of cases, $\sim 2$\% of the total flares, inaccuracies in the flare list show a start time that occurs after the peak time.  These cases were not considered in the final analysis.  The maximum $R$ value, $R_{\rm max}$, was next computed for each flare.  Figure~\ref{fig-ratio} shows an example of the X-ray flux, X-ray background level, and $R$ for a flare.  In computing $R_{\rm max}$, we required that more than one $R$ measurement must exist and that both the short and long X-ray flux $> 10^{-9}$\,W\,m$^{-2}$.  These criteria excluded 22\% of the total flares from further analyses with the $R_{\rm max}$ parameter, including near-instantaneous flares with short ($\sim 1$ minute) rise times.

Table~\ref{table-statistics} includes the average statistics for each flare class.  These statistics include the total number of flares,  time between $R_{\rm max}$ and 1-8\,\AA\,peak flux (where measurements were possible following the criteria from the previous paragraph), peak flux, background flux in the short and long band, and $R_{\rm max}$.  Since the ratio $R$ is related to coronal temperature (see, e.g.,  \citealt{1985SoPh...95..323T,1994SoPh..154..275G,1996ApJ...460.1034F,2005SoPh..227..231W,2012ApJS..202...11R}), the $R_{\rm max}$ value is related to the maximum temperature occurring during the flare, where: 
\begin{equation} T_{\rm max} = A_0 + A_1 R_{\rm max}^2 + A_3 R_{\rm max}^3 {\rm MK}. \end{equation}  $A_n$ are coefficients found in Table 2 of \citet{2005SoPh..227..231W}.  We note that this direct assumption is a simplification based on an isothermal flare. As discussed in \citet{2005SoPh..227..231W}, more extensive investigations of the multi-thermal flare properties are not possible with the two flux measurements provided with the GOES XRS but the ratio is still a useful tool in studying the overall energetics of the flares. We find that the maximum ratio is significantly higher for the strongest flares (e.g., 0.33 for X flares and 0.05 for B flares).  As shown in the table, this maximum temperature, on average, occurs before the peak in the 1--8\,\AA\,flux, in part due to the fact that the 0.5-4\,\AA\,flux peaks ahead of the 1--8\,\AA\,flux by up to 20 minutes.  The rise time from flare onset to $R_{\rm max}$ is longer for X and M flares than C and B flares, consistent with statistical results presented by e.g., \citet{2002A&A...382.1070V}.

\begin{figure}
\centering
\includegraphics[width=0.8\linewidth]{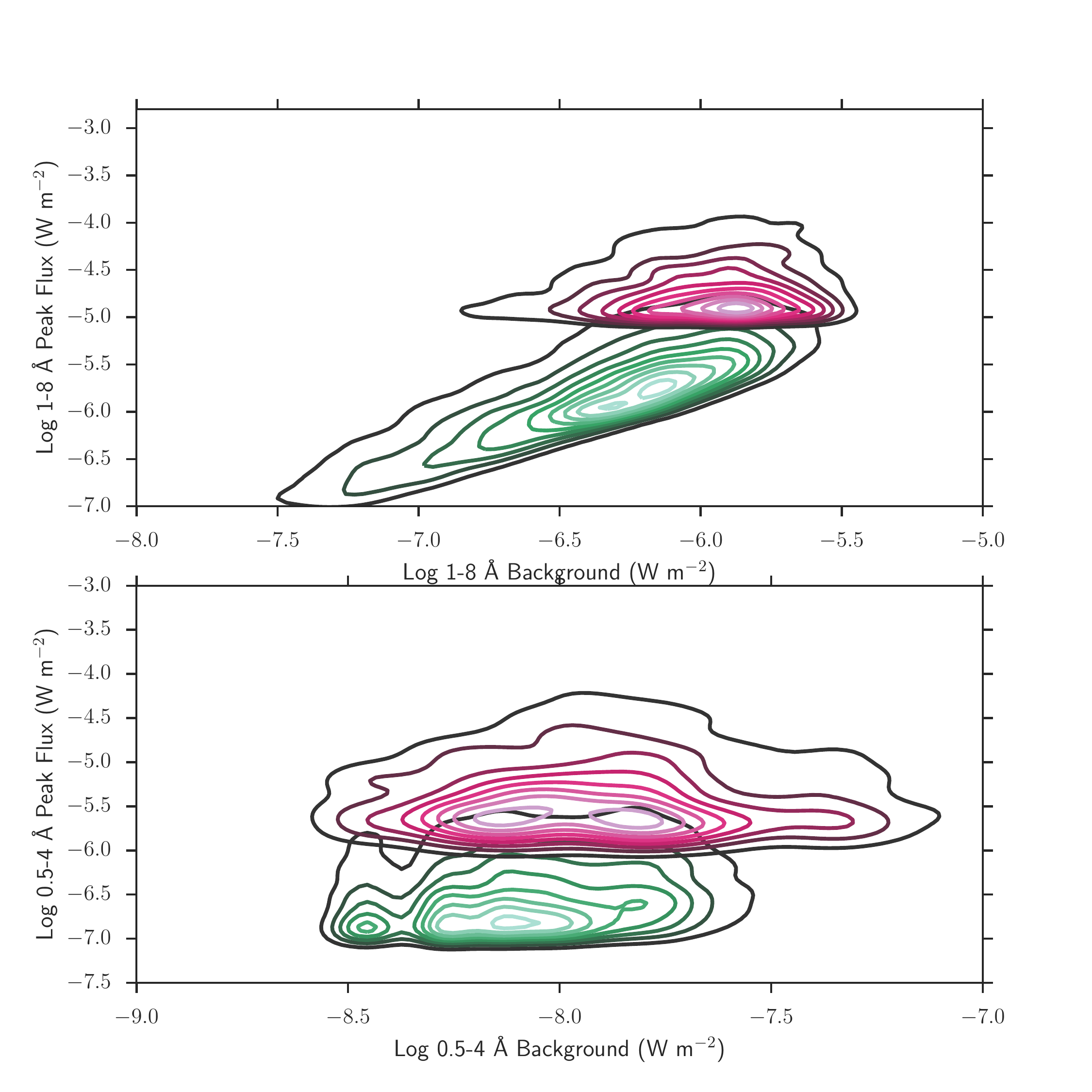} 

\caption{Kernel density estimates showing the two-dimensional probability distribution of peak flux and
background flux in both the long-wavelength (top) and short-wavelength (bottom) X-ray bands. The green contours show the density distributions for all flares, which are dominated by the more numerous C- and B-class flares, while the red contours show the density distributions for X- and M- class flares. There is no correlation between short-wavelength peak flux and background, while there is a positive correlation in the relationship between long-wavelength peak flux and background ($R^2 = 0.45$, see \S~\ref{sect-xraydata} for details). 
}\label{fig-kde}
\end{figure}

We find that the average long-wavelength background flux is higher for the stronger flares (X and M) than the weaker flares (C and B).  This is also shown in Figure~\ref{fig-kde}, with contour plots of the multivariate density estimates for the peak flux and background flux. The density estimates are created with the kernel density method, using gaussian filters, through the scientific Python (scipy) {\tt gaussian\_kde} function \citep{scott2009multivariate}. A linear correlation is found between the peak flux and background in the long-wavelength band with 
\begin{equation} \log B = (1.03 \pm 0.01) \times \log F_{{\rm peak,}~1-8~{\rm \AA}} + (0.86 \pm 0.04) \end{equation} and a correlation coefficient $R^2 = 0.45$. This is particularly evident in the full sample of flares, whose statistics are dominated by the more numerous B- and C-class flares (shown in green). However, the trend does not exist when examining the distribution of M- and X-class flares alone (shown in red, $R^2 = 0.07$). At high peak flux levels, there is no difference between the background levels (this is discussed again in the following section). Similarly, we computed density plots for the short-wavelength peak and background. Given the issue of the 0.5-4\,\AA~background being close to or below the instrumental limit, we chose to examine flares occurring from 1999-2006, a time period with higher measured backgrounds that includes the rise through fall phases of solar cycle 23 solar maximum. No correlation exists between peak flux and background in the short-wavelength band ($R^2 < 0.01$). This further illustrates that the background in the long-wavelength band and not the short-wavelength band is an appropriate observational parameter that is tied to the peak flare flux.

\section{Separation into Solar Flare Classes}\label{sect-diagram}
\begin{figure}
\centering
\includegraphics[width=0.45\linewidth]{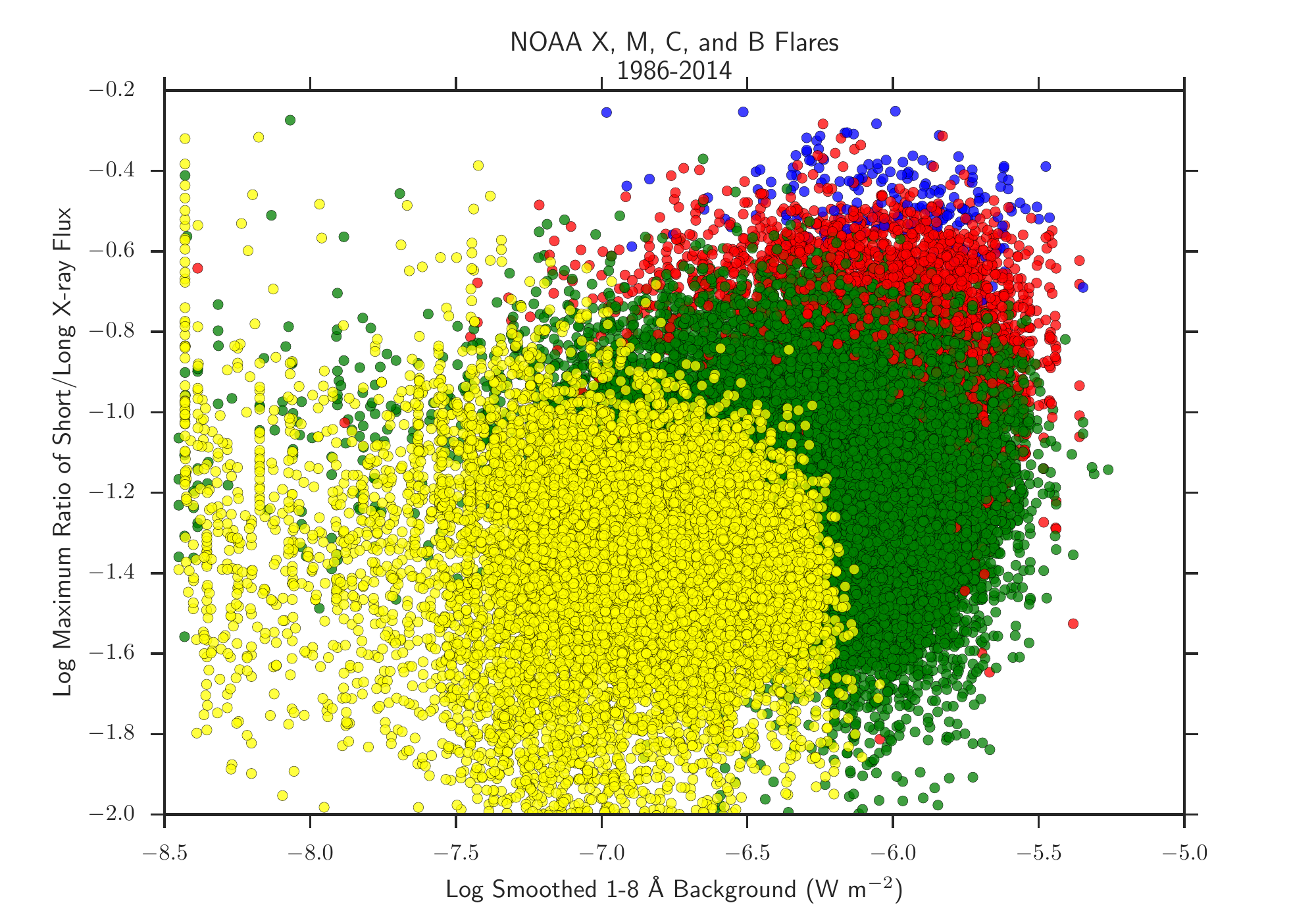}
\includegraphics[width=0.45\linewidth]{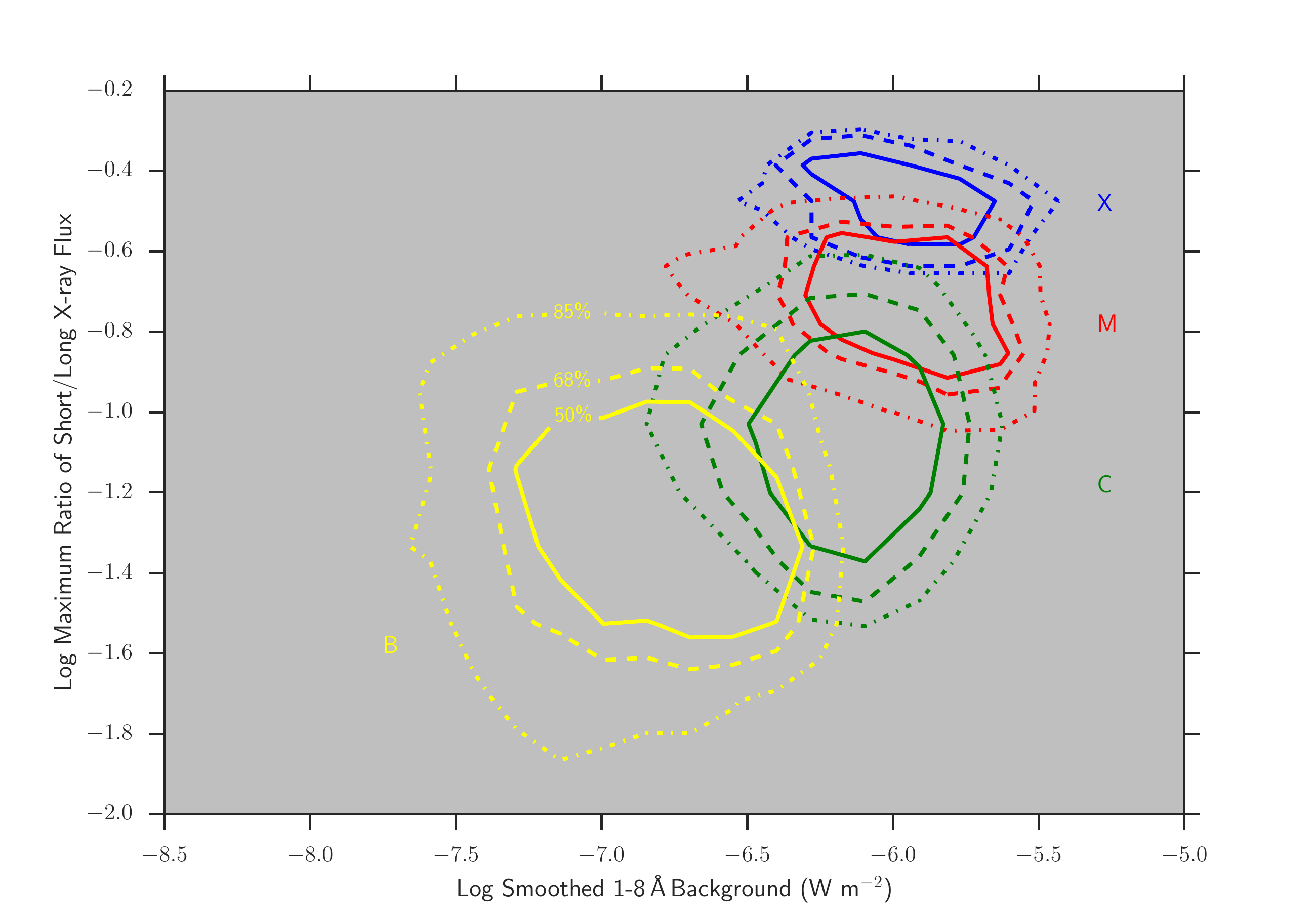}
\caption{The observational parameters of 1--8\,\AA\,non-flare X-ray background flux and the maximum ratio of the 0.5--4\,\AA/1--8\,\AA\,flux ($R_{\rm max}$) separate the NOAA flares effectively into different parameter space based on the peak flux.  The left panel shows a scatter plot of the measured parameters of the 50,000 NOAA flares (with color-coding corresponding to flare class as blue = X, red = M, green = C, and yellow = B). In the right panel, contour levels display levels enclosing 50\% (solid line), 68\% (dashed line), and 85\% (dashed dotted line) of the X- (blue), M- (red), C- (green), and B- (yellow) class flares.  Lower peak flux occurs when the background is also low.  High peak flux occurs when the background flux and $R_{\rm max}$ parameters are high.
}\label{fig-phasediagram}
\end{figure}

With the extensive database of X-ray flare properties, we investigated whether measurements based on the 1-min $GOES$ XRS observations showed properties useful for predicting the X-ray class.  Specifically, in order to build a classification model based on the properties of past flare events, we identified properties that lend themselves to the use of classification techniques by showing a separation between different flare classes in their parameter space.  The 1--8\,\AA\,non-flare background and the $R_{\rm max}$ parameters yielded such a separation, shown in Figure~\ref{fig-phasediagram}. For each of the flare classes, the contours represent the parameter space of the diagram including 50\%, 68\% (the 1-sigma contour level), and 85\% of the flares in the given class. The X-class flares occupy the right corner of the diagram, indicating high background flux and high $R_{\rm max}$.  The M-class flares share a similar range of background flux, but with lower $R_{\rm max}$ than the X-class flares.  This is also evident in Table~\ref{table-statistics}, where the average and standard deviation in $B$ is consistent between X and M flares while the average $R_{max}$ is significantly higher for the X-class.  Similarly, C-class flares share the range of backgrounds with X- and M- class flares, but have lower $R_{\rm max}$ values.  However, the B-class flares have significantly lower measurements of the background flux, with a similar range of $R_{\rm max}$ to the C-class flares. Due to the upper flux limit of B-class flares ($< 10^{-6}$\,W\,m$^{-2}$), they can not be observed when the background is high.

To further test the apparent difference in the $B$ and R$_{\rm max}$ parameters of different flare classes, we computed Kolmogorov-Smirnov statistics \citep{cite-kolmogorov, smirnov1948}.  The Kolmogorov-Smirnov two sample test determines the probability of two samples being drawn from the same distribution.  To do this, the cumulative distribution of each sample is computed and the maximum distance between the two chosen distributions is determined as the Kolmogorov-Smirnov (KS) statistic.  When the KS statistic is small and the two-tailed p-value approaches 1, the null hypothesis of both samples being drawn from the same distribution can not be rejected.  The two sample KS test was run using the scipy {\tt ks\_2samp} function \citep{cite-scipy}.  The test was run on all combinations of two flare classes (e.g., sample 1 as X-class flares and sample 2 as M-class flares) for each of the parameters $B$ and R$_{\rm max}$.  Results are given in Table~\ref{table-kstest}, including the KS statistic and the two-tailed p-value.  These statistics show that the long X-ray background is consistent with being drawn from the same population for X- and M-class flares.  To a lesser degree, the KS statistic is low ($\sim 0.3$) for comparisons between the distribution of background flux of X- and M-class flares with C- class flares, but the low p-values ($< 0.001$) indicate that these distributions are distinct. Additional comparisons between the distributions of $B$ and R$_{\rm max}$ for the flare classes result in high KS statistics (from $0.395 - 0.985$) and low p-values ($< 0.001$). This suggests that the distributions of values in the $B$-R$_{\rm max}$ parameter space are distinct. This is also evident in Figure~\ref{fig-phasediagram}, where we show that the majority of flares of each class (e.g., the 1-sigma or 68\% contour level) occupy a distinct parameter space.


This separation into flare classes hints at differences in the physical conditions of the solar corona.    The long X-ray background, $B$, is the non-flare flux level associated with active regions.  It can be construed as a proxy for the magnetic energy of the corona.  The $R_{\rm max}$ measurement is associated with coronal temperature, as well as radiative losses and emission measure (see, e.g.,  \citealt{1985SoPh...95..323T,1994SoPh..154..275G,1996ApJ...460.1034F,2005SoPh..227..231W,2012ApJS..202...11R}).  A possible explanation for the separation of flare classes in the $B$-R$_{\rm max}$ is that the built-up energy of the regions that produce the flares (measured by $B$) is directly related to the amount of energy released in the flare (measured by $R_{\rm max}$).  Since the background measurement is an average over the entire Sun and not just the flare site, we expect that a more careful analysis where $B$ is replaced by a measurement of the energy/flux of the flare site alone would reveal a tighter correlation with $R_{\rm max}$.  This, however, is a more difficult measurement to make for a real-time forecast situation.

\section{Machine Learning Classification}
To quantify the separation into flare classes shown in the $B$-R$_{\rm max}$ parameter space, we used machine learning classification techniques.  Specifically, we used the K-nearest neighbors and nearest centroid algorithms from the Python machine learning library, {\tt scikit-learn} \citep{scikit-learn}.  These classifiers build predictions using input data from a training set of data with known classes.  Our training set included X-ray flares from solar cycles 22-24.  The input parameters were the X-ray background and the maximum ratio of short to hard X-ray flux ($R_{\rm max}$), with the classes labeled as X, M, C, B.  The classifier algorithms then use the training set to predict what the class of a new flare event will be.  In \S~\ref{sect-modeldescription}, the machine learning classification techniques are described.  The machine learning algorithms applied to the input data create what is termed a model, which is a statistical model based on training data that is used to make predictions for new data sets. Results of our analysis applying our statistical models are included in \S~\ref{sect-results}.

\subsection{Statistical Model Descriptions}\label{sect-modeldescription}
For the K-nearest neighbor classifier, the parameter space of the logarithm of X-ray background vs. the logarithm of $R_{\rm max}$ is broken up into a grid.  Along each point in the grid, the number of data points of each possible class in the $k$ nearest neighboring points are counted, where $k$ is a user-defined integer.  Whichever class corresponds to the most neighboring data points is adopted in the model as the likely class of any new flares with the same values of the X-ray background and $R_{\rm max}$ at that grid point.  For our analysis, the grid size was 0.01 and the $k$ neighboring points used was 5.  As an example of computing a classification for one grid point, we consider the point where log $R_{\rm max} = -0.8$ and log of the X-ray background $= -8$.  Figure~\ref{fig-phasediagram} shows that the 5 nearest data points to the selected point include 4 B-class flares and 1 C-class flare.  Therefore, an unknown flare at the selected point would be classified as a B-class flare.  

Alternatively, the nearest centroid algorithm uses the distance from the centroid of the distribution of points in each class of the training set for predictions.  For example, we consider the classification of a point based on the distance from the centroid of the X-class flares (log $R_{\rm max} = -0.48$ and log X-ray background $= -5.92$) and M-class flares (log $R_{\rm max} = -0.72$ and log X-ray background $= -5.92$).  To classify an unknown flare with log $R_{\rm max} = -0.5$ and log X-ray background $= -5.9$, we calculate the Euclidean distance of the point to the centroid of each of the classes.  The selected point is a distance of $\sqrt{(-0.5 - -0.48)^2 + (-5.9 - - 5.92)^2} = 0.03$ from the X-class centroid and, similarly computed, 0.22 from the M-class centroid.  Since it is closer to the X-class centroid, the selected point is classified as an X-class flare.  

\begin{figure}
\centering
\includegraphics[width=0.45\linewidth]{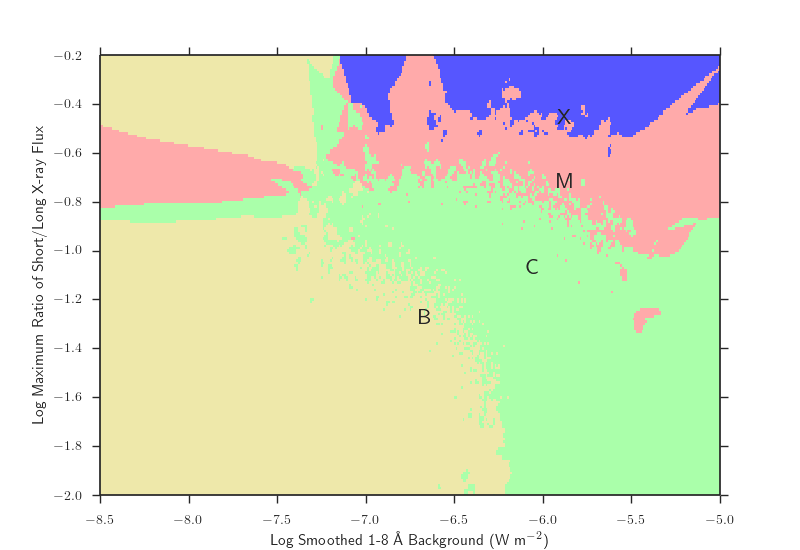}
\includegraphics[width=0.45\linewidth]{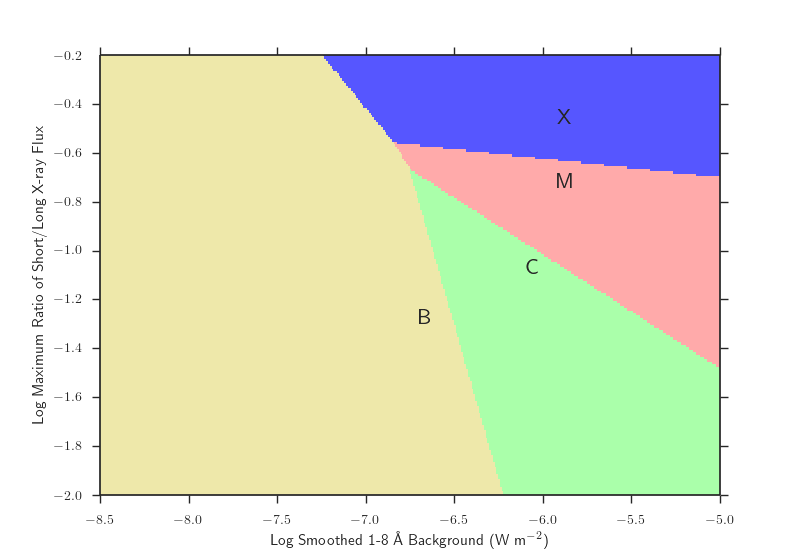}
\caption{Classification models built from the solar cycle 22 - 24 flare parameters for the K-nearest neighbor (left) and nearest centroid (right) algorithms.  The K-nearest neighbor model classifications use the classes (X, M, C, B) of the 5 nearest points in the training set to each grid point ($0.01 \times 0.01$) to predict the class of a flare with the $B$ and $R_{\rm max}$ values at that grid point.  The nearest centroid model predicts classes based on the Euclidean distance each grid point is from the centroid of $B$ and $R_{\rm max}$ for each class (X, M, C, B).  See the text for more detail on the classification algorithms.
}\label{fig-machinelearning}
\end{figure}

Since the K-nearest neighbor approach weights according to the number of points along the grid, it is more accurate for classifying C- and B- class flares, which include 4-100$\times$ more flares than the M- and X- class flares.  Meanwhile, since it is based solely on the distance from the centroid of the parameters for a given class, the nearest centroid method does a better job at classifying the X-class flares.  Using these machine learning methods, classification models were built with the $B$ and $R_{\rm max}$ measurements from flares from solar cycles 22-24.
Since there are relatively few, less than 300, X-class flares in the entire sample spanning nearly three decades, the advantage of using the cycle 22-24 data as a training set is that the resultant model will include as many X-class flares as possible. These statistical models built with the full flare dataset are shown in Figure~\ref{fig-machinelearning}.  

To create statistical models that do not include the training set data, but still include a larger number of X-class flares, we also built models using only the flare parameters from solar cycles 22 and 23. These models were then used to predict the solar cycle 24 flare classifications. A concern with using this approach is whether the flare behavior during the much weaker solar cycle 24 is different from the flares in cycles 22 and 23 that were used to create the model. To test whether there are differences in flare rate between the solar cycles, we determined the occurrence frequency rate as a function of 1-8\,\AA~peak flux during the rise to solar maximum and solar maximum phases for each of the solar cycles. The occurrence frequency distributions were fit with a power-law of the form \begin{math} {\rm N}(F_{1-8 Ang}) = N_1 \times (F_{1-8 Ang})^{-\alpha+1} \end{math}, where N is the occurrence frequency (flares rate/day), $F_{1-8 Ang}$ is the logarithm of flare peak flux (W\,m$^{-2}$), and the fit parameters include the normalization factor, $N_1$, and the power-law index, $\alpha$. We utilize the Levenberg-Marquardt least-squares minimization technique \citep{cite-levenberg,cite-marquardt} to determine the best-fit function parameters, fitting the occurrence rate where peak flux $> 10^{-6}$\,W\,m$^{-2}$.  We omit the B-class flares from these fits since the low end of the power-law distribution is near the GOES detection threshold. Goodness of fit is assessed with the $\chi^2$ statistic, defined as $\chi^2 = \Sigma({\rm data - model})^2/{\rm std}^2$, where the data are the frequency distribution values (N), the model is the power-law fit, and std is the standard deviation for each of the measurements of N. Good fits are those where $\chi^2$/dof are close to unity, where dof are the degrees of freedom or number of data points - number of free parameters that are fit.  

\begin{figure}
\centering
\includegraphics[width=0.8\linewidth]{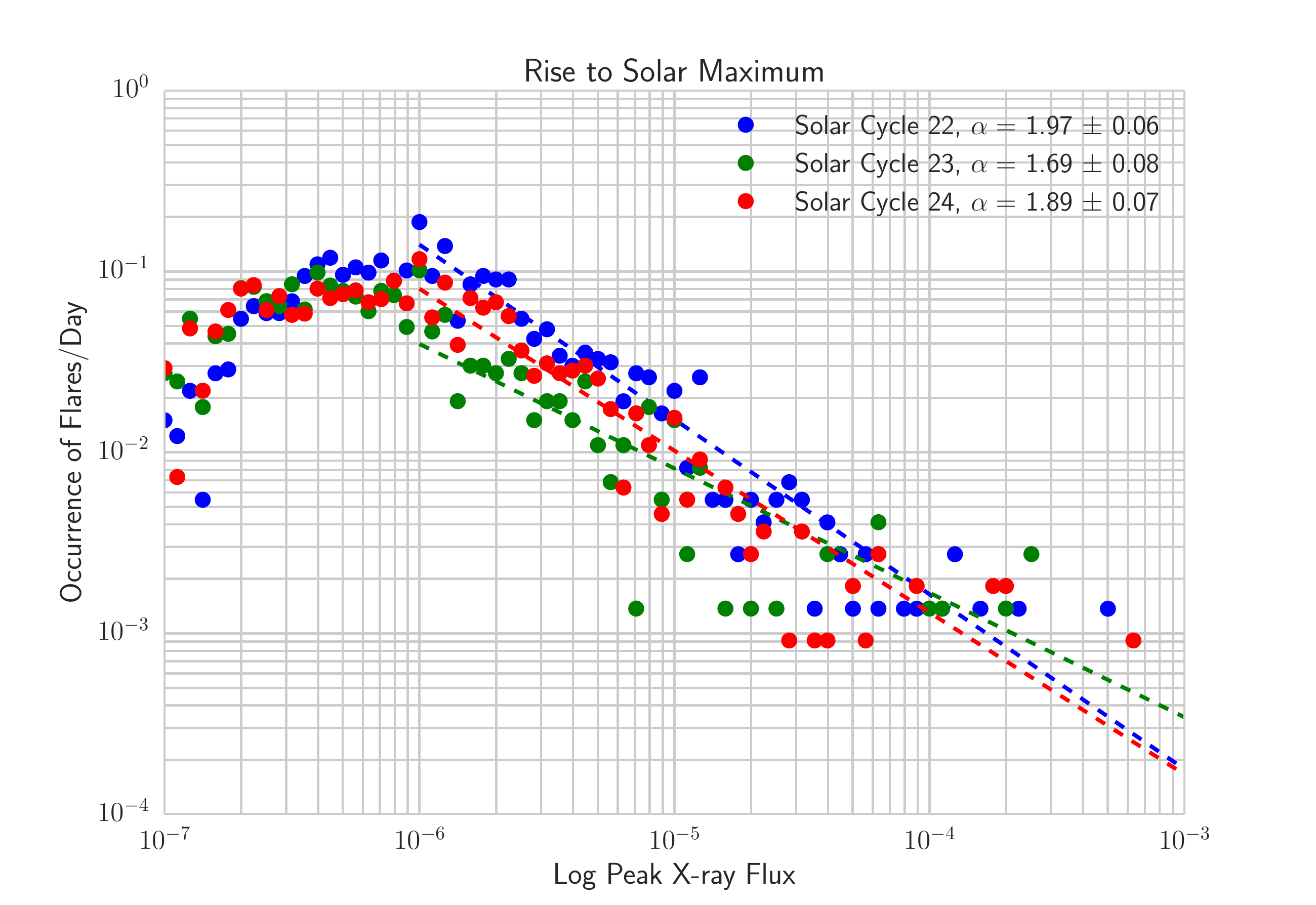}
\includegraphics[width=0.8\linewidth]{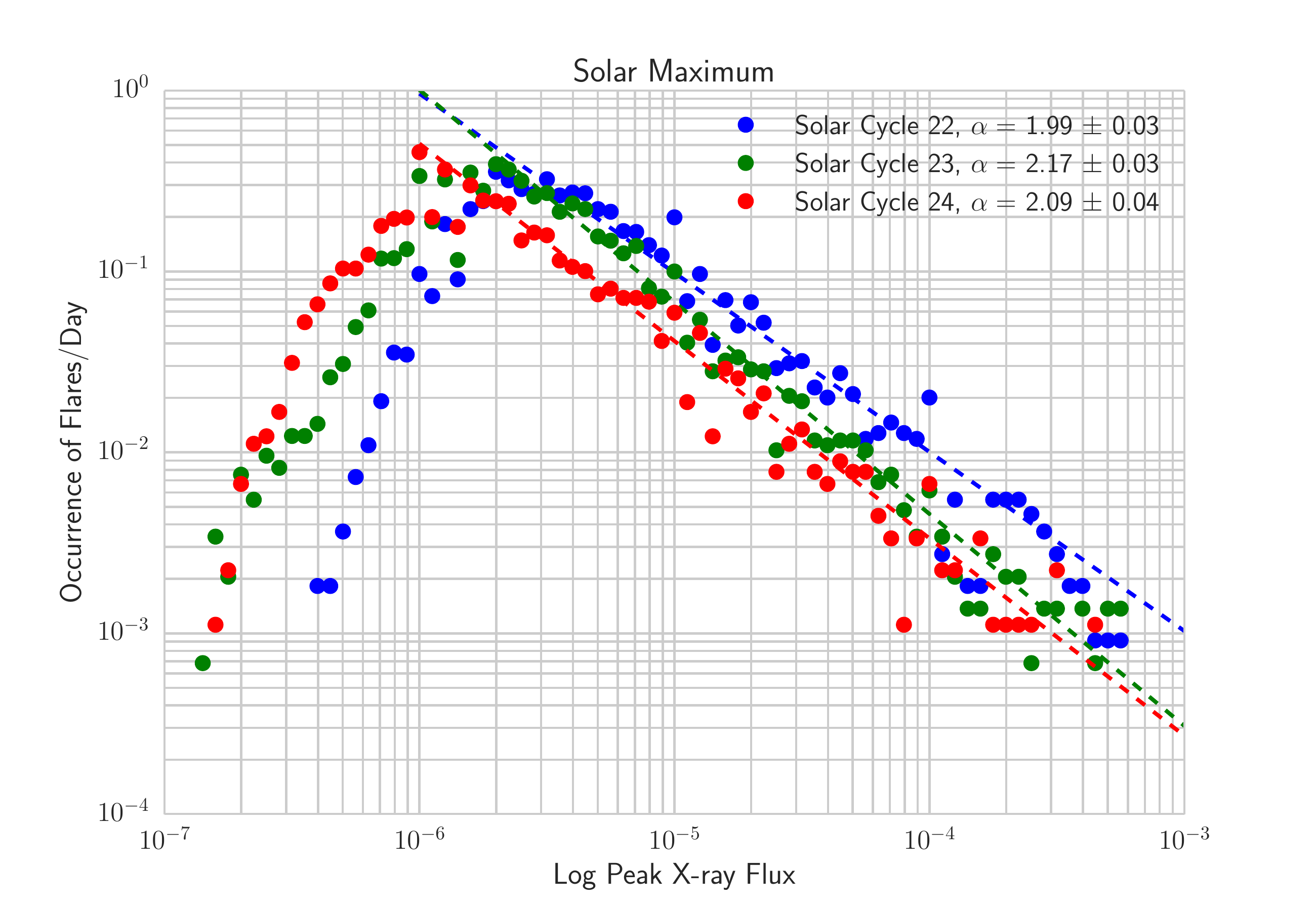}
\caption{Occurrence frequency distributions of the peak 1-8\,\AA~flux are shown for solar cycles 22, 23, and 24 both during the rise to solar maximum and during solar maximum. Flare rates are similar between solar cycles for the rise to solar maximum and during solar maximum. 
}\label{fig-occurrence}
\end{figure}

Using the solar X-ray background analysis from \citet{2041-8205-793-2-L45}, the rise phases are defined as occurring from 08/1986 - 08/1988 (solar cycle 22), 05/1996 - 05/1998 (solar cycle 23), and 12/2008 - 12/2011 (solar cycle 24) and the solar maximum phases are defined as 08/1988 - 08/1991 (solar cycle 22), 05/1999 - 05/2003 (solar cycle 23), and 12/2011 - 12/2014 (solar cycle 24, noting that this is an incomplete solar maximum phase including flares up until the end of the period examined in the paper). Plots of the frequency occurrence rates are included in Figure~\ref{fig-occurrence} and the best-fit parameters are in Table~\ref{table-occurrence}. \citet{2011SoPh..274...99A} includes power-law estimates from past analyses of frequency distributions of X-ray flares, which find $\alpha$ to range from 1.58 - 2.0. They find that the slopes change throughout the solar cycle, with flatter slopes during solar maximum. Our rates are consistent with these values. Additionally, as in \citet{2011SoPh..274...99A}, we find power-law slopes are similar during the same phase (e.g., solar maximum), but have different normalizations. Therefore, we conclude that the power-law slope of the frequency rates for flares are consistent between solar cycles and as a result we can effectively utilize the solar cycle 24 flares as an appropriate test set for classification models built with the solar cycle 22-23 flare parameters. 

\subsection{Results}\label{sect-results}
Table~\ref{table-machinelearning} presents statistics on the percent of correct identifications (PC), the number of true classifications (TC; number of flares where the correct flare class is predicted) divided by the total number of flares (N) examined.  The PC computations show the ability of the models, built with solar cycle 22-24 and solar cycle 22-23 flare parameters, to correctly classify flares from the test sets of flare parameters, solar cycle 22-24 and 24 flares.  The models built with the solar cycle 22-24 data correctly classify $\sim$ 90\% of the flares with the K-nearest neighbor model and $\sim 75$\% with the nearest centroid model.   The nearest centroid model correctly classifies 95.9\% of X-class flares with the solar cycle 22-24 tested model and correctly classifies all of the solar cycle 24 X-class flares.  The K-nearest neighbor model better predicts the M-, C-, and B- class flares, correctly classifying 66.6\% of M-class flares, 91.8\% of C-class flares, and 89.1\% of B-class flares from solar cycles 22-24.  The model does a better job of classifying the solar cycle 24 flares, with correct classifications of 80-90\% of M through B flares.  From the tests of the solar cycle 22-24 and solar cycle 22-23 built models, the performance is similar in correctly identifying solar cycle 24 flares, but the classifications are slightly better for the solar cycle 22-24 built models (by $\sim 5$\% overall) since the training set includes the solar cycle 24 flares.


Additional skill scores were computed to better quantify the results, shown in Table~\ref{table-skillscores}.  These skill scores include the probability of detection (POD), false alarm rate (FAR), Heidke skill score (HSS; see \citealt{cite-Heidke}), and true skill score (TSS; defined in 
\citealt{cite-TSSref}).  For each flare class, the following values were computed: the true classifications (TC), false null classifications (FN; number of flares in the class incorrectly predicted not to be in the flare class), false classifications (FC; number of flares not in the class incorrectly predicted to be in the flare class), and the true null classifications (TN; number of flares not in the flare class and correctly predicted not to be in the flare class).  These definitions are similar to those defined in forecasting solar energetic particle events and solar flares (recent examples include \citealt{2009SpWea...704008L} and \citealt{2012ApJ...747L..41B}).  Using these definitions:
\\
\vspace{0.2cm}
\\
\begin{math}
{\rm POD} = {\rm TC}/({\rm TC} + {\rm FN}), \\
{\rm FAR} = {\rm FC}/({\rm TC} + {\rm FC}), \\
{\rm HSS} = \frac{2 \times [({\rm TC}\times{\rm TN}) - ({\rm FN} \times {\rm FC})]}{({\rm TC} + {\rm FN})({\rm FN} + {\rm TN}) + ({\rm TC} + {\rm FC})({\rm FC} + {\rm TN})},\,{\rm and}\\
{\rm TSS} = \frac{\rm TC}{({\rm TC} + {\rm FN})} - \frac{\rm FC}{({\rm FC} + {\rm TN})}.
\end{math}
\vspace{0.2cm}

The POD values indicate that, as shown with the PC statistic, the nearest centroid model correctly predicts the X- and M-class flares better than the K-nearest neighbor model.  However, the FAR shows that the K-nearest neighbor model makes fewer false predictions of a flare incorrectly being classified in the X- or M-class.  For the X-class flares, for instance, even though all of the X-class flares are correctly classified, the FAR is high with the nearest centroid model since there are 567 non-X-class flares incorrectly classified as being in the X-class.  From a forecast stand point, this argues that both a combination of the K-nearest neighbor predictions, which have low FAR, and the nearest neighbor predictions, which have high POD, would be necessary for making solar flare predictions.  For the C- and B-class flares, the combination of low FAR and high POD for the K-nearest neighbor model proves it to be the superior model for predicting lower flux flares.

The final statistics, HSS and TSS, are commonly used skill scores with an advantage over the PC, POD, and FAR statistic in that they incorporate all of the parameters TC, FN, FC, and TN.  By using all combinations, HSS takes into account the expected number of correct identifications due to chance.  An advantage of TSS over HSS, as pointed out in \citet{2012ApJ...747L..41B}, is that the TSS does not change depending on the number of flares in the sample size.  The results of the TSS show the nearest centroid model as the best performer for the X-class flares, while the K-nearest neighbor is better for the weaker flares.  The HSS roughly agree with TSS, with the exception of the X- and M-class flares in the nearest centroid model.  The lower values of HSS are likely due to the smaller number of flares in these categories compared to the number of false predictions (for instance, there are 196 X-class flares, but 567 non-X-class flares were incorrectly predicted as X-class).

\section{Discussion}
Many of the previous studies of the properties of solar flares rely upon detailed analyses of a single or small group of flares.  However, statistical analyses of large samples of flares offer new insights into the physical properties associated with these events.  For instance, studies of the GOES X-ray light curves by \citet{2012ApJ...754..112A} tested a theoretical model (fractal-diffusive self-organized criticality) for flare generation, finding that nano flares are not likely to play a major role in flare heating, and \citet{2012ApJS..202...11R} built upon previous studies to refine peak temperature and emission measure statistics.  In this paper, we present a new way to separate solar flares into the NOAA flare classes based on a statistical analysis of the GOES X-ray observations of the $\sim$ 50,000 flares occurring from 1986 - mid-2014.  

These flare classification predictions are based upon observed X-ray properties -- the 24-hour non-flare X-ray background in the 1--8\,\AA\,band and the maximum ratio of the short to long band flux during the flare.  These parameters reveal a separation between the X-, M-, C-, and B- class flares.  The separation was quantified and verified through machine-learning algorithms and skill score statistics, applied to the solar flare parameters from solar cycles 22-24.

The R$_{max}$ parameter is related to the maximum temperature of the flare.  Using the relations and constants from \citet{2005SoPh..227..231W}, we find that the maximum temperatures of flares range from $\sim$ 16 - 49\,MK (X-class), 6 - 15 MK (M-class), 4 - 6 MK (C-class), and $\sim$ 4 - 11 MK (B-class).  These results are consistent with recent analyses, e.g., peak temperature from GOES presented in \citet{2012ApJS..202...11R}.  The maximum temperature is reached from a few minutes before to up to 25 minutes after the start of the flare.  The stronger the flare, the more potential warning time we may have to predict the peak. For example average R$_{max}$ for X-flares occurs 6.7 minutes before the peak while the time for C and B class flares is much shorter at $\sim$ 3 minutes. One potential challenge to applying this technique for real-time flare forecasting is that the predictions are made after the maximum is reached. With XRS observations of 1 minute resolution, as used in the current analysis, the short warning time is significantly decreased. In this case, however, the predictions are still useful in determining that the flare will be entering the declining phase. Also, NOAA SWPC currently provides finer time resolution XRS archival observations of a few seconds cadence. Accessing these data in real-time would greatly enhance the warning time available from our technique.

While the maximum temperature is an indicator, like the peak flux of the flare, of the energy release, the non-flare background is less straightforward to interpret.  This parameter is the integrated X-ray flux of the Earth-facing Sun, during the lowest flux period in 24-hours preceding the flare.  Integrated X-ray flux is dominated by active regions (e.g., \citealt{1996mpsa.conf....3A} showed that more than 50\% of the coronal luminosity is associated with 2\% of the solar surface). The non-flare background is therefore a measure of the active regions, or areas of enhanced coronal heating.  In past studies of X-ray imaging from SOHO and Yohkoh and full-disk magnetograms, \citet{1998ApJ...508..885F} showed that X-ray luminosity is highly correlated with the active region's unsigned magnetic flux, with L$_X \sim \Phi_{\rm tot}^{1.19}$.    \citet{2007ApJ...665.1460T} confirmed this in a study of 160 active regions and also found a strong correlation with the magnetic energy dissipation (also found earlier by \citealt{2006ApJ...646L..81A}), which they claim could be showing the importance of photospheric turbulent motions to heating of the corona above active regions.  Therefore, the X-ray non-flare background flux is an indicator of the average magnetic energy of the active regions as well as the turbulent energy of the photosphere below these active regions.  The phase-separation between strong flares and weak flares is guided by the magnetic energy available to produce flares.  Higher turbulence or stored magnetic energy leads to more energy release in the flare, measured by the maximum flare temperature.     

Since the non-flare background is measured in the 24 hours preceding the flare, it can be used for flare predictions in advance of the flare.  One way the background can be used in real-time forecasting is as a threshold for predicting when strong flares may or may not occur.  This is possible due to the large separation in the range of background values of strong flares (M and X class) versus the weakest flares (B class).  From analysis of the distributions of the background for the different flare classes, we find that at the -2$\sigma$ level (the 2.28th percentile) for M and X class flares the background is $1.6 \times 10^{-7}$\,W\,m$^{-2}$.  Therefore, there is a low probability of a background flux below this level being associated with an X or M class flare.  For C, M, and X class flares, the -2$\sigma$ level is $1.07 \times 10^{-7}$\,W\,m$^{-2}$, meaning anything lower than this background flux is unlikely to be associated with a strong flare.  

Based on these X-ray background results, during solar minimum when the background flux is low we expect no X- or M-class flares.  For instance, the results from Figure 1 of \citet{2041-8205-793-2-L45} show that solar cycle 24 had average 2-week X-ray background measurements below $1.07 \times 10^{-7}$\,W\,m$^{-2}$ for the first two years of the cycle.  This means only B-class flares would have been expected in these years (2009-2011).  From Figure 1, it is clear that relatively few strong flares had occurred during this time.  During these two years, the NOAA list records 816 flares, including no X-class flares, 13 M-class flares (1.6\% of flares), 103 C-class flares (12.6\% of flares), and 700 B-class flares (85.8\% of flares).  Since this is a rough estimate of the background based on 2-week averages, we expect that the background occasionally rose above this low threshold, accounting for the small number of M-class flares observed during the last solar minimum.  These flare rates during the beginning of solar cycle 24 are similar to those in the first two years of solar cycles 22 and 23 for X- and M- class flares, but, likely due to the higher backgrounds in cycles 22 and 23, there are more C-class and fewer B-class flares in cycles 22 and 23. For comparison, the beginning of solar cycle 22 had 1460 flares (2 X-class, 41 M-class, 410 C-class, and 1007 B-class flares from 1986-1988) and solar cycle 23 had 914 flares (4 X-class, 18 M-class, 262 C-class, and 630 B-class flares from 1996-1998).

Additional investigation into the relationship between temperature and non-flare X-ray background will lead to refinement of our flare predictions.  Future work will investigate how the separation is affected by choice of the non-flare background.  For instance, for forecasting purposes the goal is to have a longer lead time before the flare occurs.  Different binning periods for non-flare background can be tested to determine how much lead time is possible while still preserving the phase-separation between weak and strong flares.  We will also look into increasing the warning time for flare class predictions, by determining whether other observables help us determine what the R$_{max}$/peak value will be further in advance.  One such possible path is through determining characteristic flare shapes for the rise time, which can be used at the start of the flare to predict when the maximum will occur.


\acknowledgements
The GOES XRS data used in this paper are available at: 

\tt{http://www.ngdc.noaa.gov/stp/satellite/goes/dataaccess.html}. 
The NOAA X-ray Flare Lists were downloaded from NOAA NGDC through the FTP site linked from here: 

\tt{http://www.ngdc.noaa.gov/nndc}. 

The monthly sunspot number was obtained from NASA Marshall Space Flight CenterÕs compilation available here: 

\tt{http://solarscience.msfc.nasa.gov/greenwch/spot\_num.txt}.

KSB expresses gratitude to AFOSR for supporting the AFRL Task on Flares and CMEs. LMW is supported by AFRL Contract FA8718-05-C-0036.  The authors thank Doug Biesecker (NOAA SWPC) for useful discussion during the 2014 SHINE conference that led to investigating the $R_{\rm max}$ ratio, which became a focal point of this paper.


\end{article}

\clearpage

\begin{figure}
\centering
\includegraphics[width=0.45\linewidth]{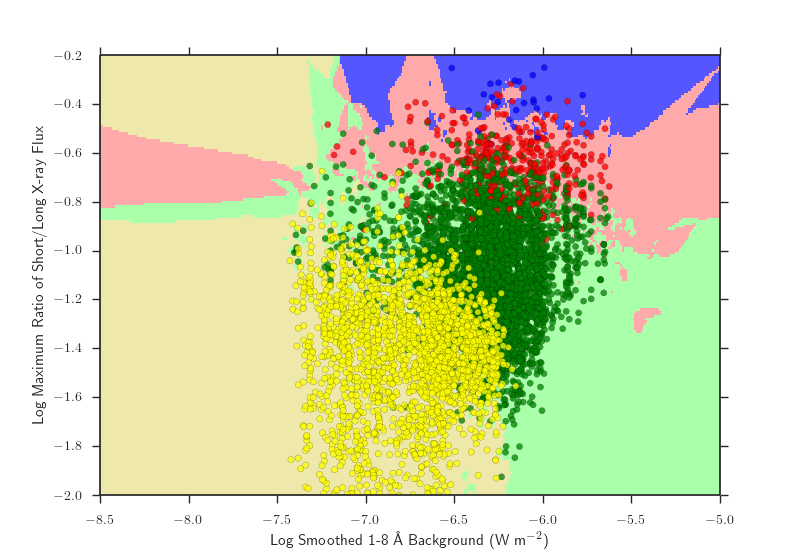}
\includegraphics[width=0.45\linewidth]{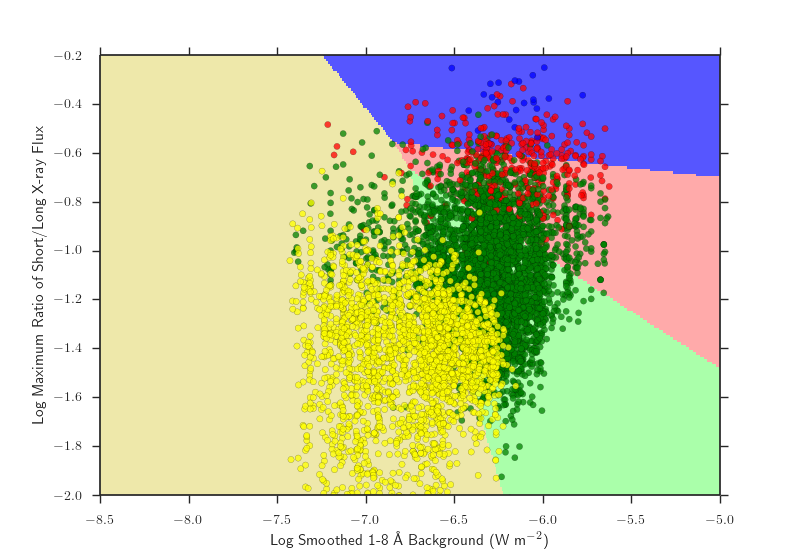}
\caption{Results from classification models built with the solar cycle 22 and 23 flare parameters for the K-nearest neighbor (left) and nearest centroid (right) algorithms, applied to the solar cycle 24 data (points, color-coded as in Figure~\ref{fig-phasediagram}).  The background shading indicates the model predicted class (color-coded as in Figure~\ref{fig-machinelearning}).  The K-nearest neighbor algorithm correctly classifies 83\% of the flares.  However, the nearest centroid algorithm classifies the highest peak flux flares (X) with 100\% accuracy compared to the 74\% accuracy from the K-nearest neighbor model.
}\label{fig-cycle24}
\end{figure}

\clearpage
\begin{table}[ht]
\caption{X-ray Flare Statistics.}\label{table-statistics}
\vspace{0.25cm}
\begin{center}
\begin{tabular}{c c c c c c c}
\hline\hline
Class & N & $<T(P - R_{\rm max})>$ & $<P_{1-8\AA}>$ & $<B_{0.5-4\AA}>$ &$<B_{1-8\AA}>$ & $<R_{\rm max}>$ \\
&	& min &  10$^{-7}$ W\,m$^{-2}$ &  10$^{-8}$ W\,m$^{-2}$ & 10$^{-6}$ W\,m$^{-2}$ & \\
\hline
X	&290 &	6.7	$\pm$ 13.2 &	2400 $\pm$ 1700 & 1.9 $\pm$ 2.1 & 1.2 $\pm$	0.7 & 0.33 $\pm$ 0.08 \\
M	& 3742 & 4.8 $\pm$ 9.9 &	240 $\pm$ 180 &	1.6 $\pm$ 1.7 & 1.2	$\pm$ 0.7 & 0.18 $\pm$ 0.06 \\
C	& 28803 & 3.0	$\pm$ 7.0 &	31 $\pm$ 20 &	0.8 $\pm$	1.0 &	 0.8	$\pm$ 0.5 & 0.08 $\pm$ 0.04 \\
B	& 15751 & 2.7	$\pm$ 6.3 &	4.9 $\pm$	2.6 &	 0.2 $\pm$ 0.2 & 0.2	$\pm$ 0.1 & 0.05 $\pm$ 0.15 \\
\hline
\end{tabular}

The statistics include number of flares (N), average and standard deviation of the time between the 1-8\,\AA\,Peak and $R_{\rm max}$ ($<T(P - R_{\rm max})>$), average and standard deviation of the peak flux ($<P_{1-8\AA}>$), average and standard deviation of the background 0.5-4\,\AA\,flux ($<B_{0.5-4\AA}>$), average and standard deviation of the background 1-8\,\AA\,flux ($<B_{1-8\AA}>$), and average and standard deviation of the maximum ratio of 0.5-4\,\AA\,/1-8\,\AA ($<R_{\rm max}>$), for each flare class.

\end{center}
\end{table}

\begin{table}[ht]
\caption{Kolmogorov-Smirnov Statistics for Two Sample Comparisons.}\label{table-kstest}
\vspace{0.25cm}
\begin{center}
\begin{tabular}{|c| c c |c c|}
\hline\hline
 & \multicolumn{2}{c}{$B$} & \multicolumn{2}{c}{R$_{\rm max}$} \\
{\bf Classes} & KS & p & KS & p \\
\hline
X, M & 0.056 & 0.600 & 0.776 & 0.000 \\
X, C & 0.298 & 0.000 & 0.976 & 0.000 \\
X, B & 0.857 & 0.000 & 0.985 & 0.000 \\
M, C & 0.283 & 0.000 & 0.706 & 0.000 \\
M, B & 0.806 & 0.000 & 0.888 & 0.000 \\
C, B & 0.695 & 0.000 & 0.395 & 0.000 \\
\hline
\end{tabular}

The K-S statistic (KS) and probability value (p) are listed for tests on the distributions of the long X-ray background and R$_{\rm max}$ for all combinations of comparisons of two X-ray flare classes.
\end{center}
\end{table}

\begin{table}[ht]
\caption{Best-fit Parameters for Power-Law Fits to the Occurrence Flare Rates.}\label{table-occurrence}
\vspace{0.25cm}
\begin{center}
\begin{tabular}{|c| c c c |}
\hline\hline
{\bf Solar Cycle Phase} & $N_1$ & $\alpha$ & $\chi^2$/dof  \\
\hline
22 Rise & -6.65 $\pm 0.29$ & 1.97 $\pm 0.06$ & 33.2/35 \\
23 Rise & -5.53 $\pm 0.40$ & 1.69 $\pm 0.08$ & 40.8/23 \\
24 Rise & -6.46 $\pm 0.34$ & 1.89 $\pm 0.07$ & 34.7/29 \\
\hline
22 Maximum & -5.96 $\pm 0.15$ & 1.99 $\pm 0.03$ & 74.7/47 \\
23 Maximum & -7.03 $\pm 0.16$ & 2.17 $\pm 0.03$ & 28.7/46 \\
24 Maximum & -6.84 $\pm 0.20$ & 2.09 $\pm 0.01$ & 28.6/41 \\
\hline
\end{tabular}

Occurrence frequency distribution for the rise phase (Rise, from the beginning of the solar cycle towards maximum) and during solar maximum (Maximum) were fit with a power-law model (see \S~\ref{sect-modeldescription} for details). The best-fit value and errors are shown for the normalization factor ($N_1$, the normalization factor for the logarithm of the 1-8\,\AA~peak flux in W\,m$^{-2}$) and the power-law index ($\alpha$). The goodness of fit is assessed with the reduced $\chi^2$ statistic ($\chi^2$ divided by the degrees of freedom, dof). 
\end{center}
\end{table}

\clearpage
\begin{table}[ht]
\caption{Classification Model Statistics.}\label{table-machinelearning}
\vspace{0.25cm}
\begin{center}
\begin{tabular}{|c | c c c | c c c | c c c |}
\hline\hline
Class & N & KNN $PC$ & NC $PC$ & N$$ & KNN $PC$ & NC $PC$  & N & KNN $PC$ & NC $PC$ \\
& & & & & & & & & \\
\hline
{\it Model} & 22-24 & 22-24 & 22-24 & 22-24 & 22-24 & 22-24 & 22-23 & 22-23 & 22-23 \\
{\it Test Set} & 22-24 & 22-24 & 22-24 & 24 & 24 & 24 & 24 & 24 & 24 \\
\hline
All & 39391 & 88.9 & 75.0 & 7032 & 88.9 & 73.7 & 7032 & 83.4 & 73.0\\
X	& 196 &	59.2 & 95.9 & 23 & 73.9 & 100 & 23 & 73.9 & 100 \\
M	& 2964 & 66.6 & 70.5 & 349 & 80.5 & 53.6 & 349 & 75.6 & 51.0 \\
C	& 23425 & 91.8 & 72.9 & 3987 & 90.6 & 73.7 & 3987 & 85.5 & 72.8 \\
B	& 12806 & 89.1 & 79.7 &2673 & 87.4 & 76.2 & 2673 & 81.4 & 76.0 \\
\hline
\end{tabular}

Models were built using the K-nearest neighbor (KNN) and nearest centroid (NC) methods, using the solar cycle 22-24 flare parameters.  The number of flares in each category (N), along with the percent of correct classifications are shown for the model built and applied to the flare data (e.g., KNN $PC$ is the percent correct for the KNN model).  The solar cycles used to build each of the models are listed in the row labeled {\it Model} and the listed statistics are for testing the model on the solar cycles listed in the row labeled {\it Test Set}.
\end{center}
\end{table}

\begin{table}[ht]
\caption{Skill Scores.}\label{table-skillscores}
\vspace{0.25cm}
\begin{center}
\begin{tabular}{|c c | c c c c | c c c c |}
\hline\hline
& & & KNN & Model & & & NC & Model&\\
Class & N & POD & FAR & HSS & TSS & POD & FAR & HSS & TSS \\
\hline
X	& 196 &	0.59 & 0.28 & 0.65 & 0.59 & 	0.96 & 0.75 &  0.39 & 0.94 \\
M	& 2964 & 0.67 & 0.22 & 0.70 & 0.65 & 	0.71 & 0.66 & 0.40 & 0.60 \\
C	& 23425 & 0.92 & 0.10 & 0.78 & 0.77 & 	0.73 & 0.14 & 0.53 & 0.55 \\
B	& 12806 & 0.89 & 0.11 & 0.84 & 0.84 & 	0.80 & 0.19 & 0.71 & 0.71 \\
\hline
\end{tabular}

The number of flares (N), probability of detection (POD), false alarm rate (FAR), Heidke skill score (HSS), and true skill score (TSS) presented for the K-nearest neighbor (KNN) and nearest centroid (NC) models built with the solar cycle 22-24 flare parameters and used to classify the same sample of flares.
\end{center}
\end{table}

\end{document}